\documentclass[apj]{emulateapj}
\usepackage{amsmath}
\usepackage{natbib}
\usepackage{graphicx}

\newcommand{\beq}{\begin{equation}}
\newcommand{\eeq}{\end{equation}}
\newcommand{\bea}{\begin{array}}
\newcommand{\eea}{\end{array}}

\begin{document}

\title{Where to Find Habitable "Earths" in Circumbinary Systems}
\author{Hui-Gen Liu$^*$, Hui Zhang, Ji-Lin Zhou}
\affil{School of Astronomy and Space Science \&  Key Laboratory of Modern Astronomy and Astrophysics in Ministry of Education, Nanjing
University, Nanjing, 210093, China \\
($^*$Email for correspondence: huigen@nju.edu.cn)}

\begin{abstract}
Hitherto, six P-type planets are found around five binary systems, i.e. Kepler-16 b, 34 b, 35 b, 38 b, 47 b, c, which are all Neptune
or Jupiter-like planets. The stability of planets and the habitable zones are influenced by the gravitational and radiative
perturbations of binary companions. In this Letter, we check the stability of an additional habitable Earth-mass planet in each
system. Based on our simulations in 10 Myr, a habitable "Earth" is hardly stable in Kepler-16 while a stable "Earth" in Kepler-47
close to the boundaries of the habitable zone is possible. In contrast, Kepler-34, 35 and 38 seem to have high probabilities of being
able to tolerate a stable "Earth" in their habitable zones. The affects of transit time variations are quite small due to the small
mass of an undetected "Earth", except that of Kepler-16 b. With a time precision of $10^{-3}$ day($\sim 88$s), an "Earth" in the
corotational resonance with Kepler-16 b can be detected in 3 years, while habitable "Earths" in Kepler-34 and 38 systems can be
detected in 10 years. Habitable "Earths" in Kepler-35 and 47 are not likely to be detected in 10 years under this precision.

\end{abstract}

\section{Introduction}

Searches for Earth-like exoplanets, especially the Kepler mission, make great contributions to our knowledge about terrestrial
planets. To date, about 66 Earth-like($<10 M_\oplus$) planets are confirmed\footnote{Data from http://exoplanets.org} and more than
1100 candidates with radii $R_{\rm p}<2R_\oplus$ are detected.\footnote{Data from http://www.nasa.gov}
Considering these Earth-like planets with minimum masses $<10M_\oplus$, only few of them have a moderate equilibrium temperature and
are considered as habitable planets \citep{Kopparapu13}, e.g., HD40307 g \citep{Tuomi12} and GJ667C c \citep{Del12}. Meanwhile, 11
candidates with radii $<2 R_\oplus$ are in the habitable zone \citep{Gaidos13}. Habitable Earth-like planets are likely to stay in
more compacted multi-planet systems. For example, HD40307 contains six planets and GJ667C has at least two planets.

Planets in habitable zone (hereafter HZ) must have a moderate temperature to maintain liquid water on the surface and have a
possibility of sustaining life \citep{Kast93,Jones01}. The temperature of a planet is usually considered via several conditions
\citep{Gaidos13}, e.g., planetary albedo, orbital eccentricity and obliquity \citep{WP03}, components of the planetary atmosphere
\citep{Pie11}, etc. In   single star systems, the HZ is fixed due to the stable flux of the host star in main sequences. However, the
HZ in binary systems is different due to the radiative perturbation of the binary companion. \citet{Eggl12} provided an analytic
method to estimate the HZs for S-type planetary orbits, and pointed out that HZs are strongly affected by the eccentricity of the
binary. \citet{KH13} studied HZs for P-type planetary orbits, which would oscillate with time due to the different orbital phases of
the binary stars. The stabilities of planets in binary systems are also different from those in single
star systems \citep{HW99}. 
 Considering a multi-planet system, planets with equal masses are stable in single star systems if their
separations exceed a critical value $\sim 7-10$   Hill radii \citep{Cham96,Zhou07}. Adding the perturbation of a binary companion, the
stabilities of planets around the binary are queried.

In this Letter, we focus on the five circum-binary systems: Kepler-16 \citep{Doyle11}, 34, 35 \citep{Welsh12}, 38 \citep{Oroze12b} and
47 \citep{Oroze12a}. All six planets detected are Jupiter-like or Neptune-like planets(see Table 1), which probably have a thick atmosphere
and may not form a solid surface, thus they are not suitable for life.  
As noted by \citet{Gong13}, the formation of habitable Earth-like planets in circumbinary
systems is possible.  Here we assume that an additional "Earth" formed in these systems  
to  obtain the tolerant regions of stable habitable "Earths". We also simulate the transit time variations (hereafter TTVs) of the
existing planets with and without an additional "Earth". The differences between these two TTVs may help us to predict the existence
of an undetected "Earth" through further observations \citep{Agol05,Lith12}.


\section{Addition-"Earth" Model}
Researchers usually estimate the boundaries of HZ via the maximum greenhouse effects provided by a CO$_2$ atmosphere and via runaway
greenhouse effect (i.e., loss of water, \citep{Kast93,Under03}). As \citet{Kopparapu13} pointed out most recently, the stellar fluxes
(with unit 1368 W m$^{-2}$) at the boundaries of HZ are \beq S_{\rm in}=S_{\rm
eff\odot}+aT_{\ast}+bT^2_{\ast}+cT^3_{\ast}+dT^4_{\ast}, \label{Sin}\eeq \beq S_{\rm out}=S_{\rm
eff\odot}+aT_{\ast}+bT^2_{\ast}+cT^3_{\ast}+dT^4_{\ast}, \label{Sout} \eeq where $T_{\ast}=T_{\rm eff}-5780$K and the value of $S_{\rm
eff\odot}, a, b, c$ and $d$ are listed in \citet{Kopparapu13}, see Tab.3 therein. In single star systems, $T_{\rm eff}$ is the
effective temperature of the star. In binary systems, the equivalent $T_{\rm eff}$ of the binary are calculated by Wien's displacement
law after combining the stellar spectral radiances of binary stars (see Section 3.2 in \citet{KH13}).

The equilibrium temperature of a planet $T_p$ is determined in two steps. (1)Estimate the combined flux on the planet $S$ by
$S=S_1+S_2$, where $S_i=\sigma(1-\alpha) (R_i/r_i)^2 T^4_{\rm eff,i}$, ($i$=1,2), 
where $\sigma$ is the Boltzmann constant, $\alpha$ is the albedo of the planets, $R_i$ is the radius of star $i$, $r_i$ is the
distance between the planet and star $i$ and $T_{\rm eff,i}$ is the effective temperature of star $i$. (2)Estimate $T_p$ of the planet
by energy equilibrium: $\pi r^2_p \times S=4\pi\sigma r^2_p T^4_p$, therefore we obtain  \beq T_p=(S/4\sigma)^{1/4}. \label{Tp} \eeq
Substituting Equations (\ref{Sin}) and (\ref{Sout}) into Equations (\ref{Tp}), we get the temperatures at the HZ boundary, $T_{\rm
in}$ and $T_{\rm out}$. Note that $T_{\rm in}$ and $T_{\rm out}$ are independent of albedo $\alpha$, but the location of the HZ
depends on $\alpha$ in the expression of $S_1$ and $S_2$.



The configurations of the  five observed circumbinary systems are listed in Table 1 according to observations. We simulate 100 cases
for each system. In each case, we put an additional "Earth" with a mass$=1 M_\oplus$ and let the "Earth" evolve under the gravities of
both binaries and the existing planets. All these "Earths" are put in circular orbits with an inclination=1$^o$ initially; the phase
angles of the "Earths" are chosen randomly. The initial locations of these "Earths" in each systems are interpreted in Section 3.
Using the MERCURY package for $N$-body simulations \citep{Cham99}, we estimate their stabilities and habitats based on our
simulations. Due to the variation of  $T_p$ , only those "Earths" with $T_p(t)\in[T_{\rm out},T_{\rm in}]$ are considered habitable in
this work.

\section{Results in Five Systems}
\subsection{Kepler-16}
The equivalent temperature of binaries in the Kepler-16 system is $T_{\rm eff}=4307.8$K,
and the HZ with temperatures [200.2K,271.6K](Figure 1(a)) is located from 0.4 to 0.75 AU approximately. The locations of HZ boundaries
change little with time because of the small contribution to the combined flux from the much cooler companion star. Note that the locations of HZ boundaries here are averaged by different azimuths.  
According to \citet{Qua12}, "Earths" inside 0.66 AU ($\approx 222$K) are unstable in the system. 
The initial locations of our 100 "Earths" are chosen between [173K, 222K] randomly.

The red/blue circles represent the initial locations of unstable/stable "Earths" during our simulations (the same notations are used
in all figures hereafter). After 10 Myr evolution, more than 90\% of "Earths" are ejected mainly due to the perturbations of Kepler-16
b, and only seven "Earths" survive outside the HZ. The orbits of other detected giant planets are also allowed to evolve under the
gravities of planets and stars in our simulations. However, due to the large mass of Kepler-16 b, if the "Earth" is stable, the orbits
of the presently observed planets are nearly unchanged. The same is true for other systems in Section 3. Finally, we obtain a null
"Earth" in the HZ of the Kepler-16 systems. As shown in Figure 1(b), we find two stable regions: $r\sim$0.72 and 0.95AU. The inner two
are in the corotational resonance with Kepler-16 b. Averaged by the time, their mean $T_p\approx$207.5 and 207.4 K, with variations
about 19.7 and 16.2 K. Therefore, these two "Earths" could be in the HZ only by considering some other different atmosphere models.

We set the albedo $\alpha=0$ in Figure 1. If we change the albedo to $\alpha$=0.3 comparable with the Earth and a higher value 0.5 to
model metal-rich planets with a smooth surface, the equilibrium temperature of planets $T_p$ becomes 0.915 and 0.841 times of that
with $\alpha=0$. The two planets in the same orbit with Kepler-16 b will obtain a lower $T_p\sim189.8$K for $\alpha=0.3$ and
$T_p\sim174.5$K for $\alpha=0.5$, which no longer seems likely to be habitable.


\subsection{Kepler-34}
The $T_{\rm eff}$ of Kepler-34 binary is 5892 K, and an "Earth" with $T_p$ in [214.2K, 282.9K] is habitable(Figure 2(a)). The HZ is
about 1.6-2.8 AU according to the fitting curve of $T_p$ in Figure 2(b) (hereafter the same). The HZ boundaries vary with time
more obviously than others because of the large eccentricity of binary stars and their similar $T_{\rm eff}$. However, the variation
of the inner HZ boundary is still very small($<0.02$ AU) since the HZ is far away from the binary compared with the separation of
binaries. The initial locations of our 100 "Earths" are chosen between [173K, 570K] randomly.

The green filled circles represent the initial locations of stable habitable "Earths" (hereafter the same). Comparing with Kepler-16,
the HZ in Kepler-34 is more stable for an additional "Earth". As shown in Figure 2(a), 74 "Earths" far away from Kepler-34 b easily
survive after 10 Myr including 32 (75\%) of the initial habitable "Earths" that remained in the HZ throughout that time. Figure 2(b)
shows the locations of the 74 stable "Earths" and their $T_p$. The $T_p(r)$ is similar to that in single star systems as a power law:
$T_p=359.1 \rm{K}(r/1 \rm{AU})^{-1/2}$. We find a stable boundary of about 1.5AU, outside which "Earths" are stable. The mean distance
of the outer HZ boundary is about 2.8 AU. Thus, the Kepler-34 system is likely to tolerate a habitable "Earth" between 1.6 and 2.8 AU
when $\alpha=0$. Due to the perturbation of Kepler-34 b, "Earths" close to it have a larger variations of $T_p$.

Changing the albedo $\alpha$ to 0.3 and 0.5, the location of the HZ changed to an inner region, thus the outer boundary of HZ becomes
2.34 AU for $\alpha=0.3$ and 1.98 AU for $\alpha=0.5$. Overlapping with the stable boundary $>$1.5 AU, we conclude that the tolerant
region of habitable Earths in the Kepler-34 system becomes: 1.6-2.34 AU with $\alpha=0.3$ and 1.6-1.98 AU with $\alpha=0.5$(see Table
2).

\subsection{Kepler-35}
The equivalent temperature of binaries in Kepler-35 is $T_{\rm eff}=5435.5$ K, and the HZ is from 1.2 to 2.0 AU with temperature
[210.0K, 278.9K] (Figures 2(c) and (d)). Both the separation and eccentricity of the binary orbit are smaller than that of Kepler-34,
therefore, the HZ boundaries are less influenced($<0.01$ AU) by the motion of binaries with time. The same effect occurs in Kepler-38
and 47, in which the HZ boundaries are nearly circular. The initial 100 "Earths" are choosen between [173K, 520K] randomly.

Figure 2(c) indicates that 15 "Earths" in the inner region are unstable while other 85 "Earths" outside still orbit the binary stars
after 10 Myr, including 36 ($\sim$90\%) of the initial habitable "Earths" that remained in the HZ.
The $T_p$ of the 85 stable "Earths" are shown in Figure 2(d), well fitted by a power law $T_p=298.8 \rm{K} (r/1 \rm{AU})^{-1/2}$. We
find that the tolerant region of a stable habitable "Earth" with $\alpha=0$ in the Kepler-35 system is about 1.2-2.0 AU and the
variations of $T_p$ of "Earths" in the HZ are all less than 8 K.

Taking $\alpha$=0.3 and 0.5, the location of the HZ will move to an inner region, i.e., 1.0-1.67 AU for $\alpha=0.3$ and 0.85-1.41 AU
for $\alpha=0.5$. Because smaller HZ overlaps with part of the unstable region, a smaller fraction of surviving "Earths" remain in the
HZ in comparison with $\alpha=0$(see Table 2).

\subsection{Kepler-38}
With a radius of $0.3964R_J$, a mass of $\sim 0.078M_J$ for  Kepler-38 b  is estimated  by the density of Neptune in our simulations.
The equivalent temperature of binaries is $T_{\rm eff}=5496.7$K, and the HZ boundary is [210.5 K, 279.4 K], about 1.6-2.9 AU (Figures
2(e) and (f)). Here The initial 100 "Earths" are chosen randomly from 1.2-4.3 AU with temperature [173K, 350K].

Due to the tight binary orbit, the large distance between "Earths" and the small Kepler-38 b, "Earths" in this system are all stable
after 10 Myr (Fig.2e). Finally, 38 ($>$90\%) of the initial habitable "Earths" remaining in the HZ. Fig.2f shows the fitting curve
$T_p=359.7 \rm{K}(r/1 \rm{AU})^{-1/2}$ and indicates that the boundary of the HZ is from 1.6 to 2.9 AU with $\alpha=0$. The mean
variations of $T_p$ of "Earth" in the HZ are all less than 10K.

The HZ becomes 1.34-2.43 AU for $\alpha=0.3$ and 1.13-2.05 AU for $\alpha=0.5$. Since all 100 "Earths" in the Kepler-38 system are
stable, most "Earths" in the HZ initially are still habitable due to our simulations while changing the albedo to 0.3 or 0.5(see Table
2).

\subsection{Kepler-47}
Using the density of Neptune, we estimate the masses of Kepler-47 b, c as $0.0248M_J$ and $0.0933M_J$, respectively. Their
eccentricities are only limited as $<0.035$ and $<0.411$, respectively. We choose a mean value for each planet, i.e.,
$0.035/2\sim0.017$ for 47b and $0.411/2\sim0.205$ for 47c in our simulations. The equivalent temperature of binaries is $T_{\rm
eff}=5501.5$K, and the HZ boundary becomes [210.6 K, 279.4 K], about 0.9-1.6 AU (Figure 2(g),(h)). The initial 100 "Earths" are chosen
randomly between [173 K, 550 K], about 0.2-2.6 AU.

After 10 Myr evolution, only 66 "Earths" survive and 7 of them are habitable, as shown in Figure 2(g). Since Kepler-47 c is located in
the HZ with eccentricity 0.2, it sweeps out a region of 0.8-1.2 AU, as shown in Figure 2(h).
Similar to the Kepler-16 system, Kepler-47 c also captures an "Earth" in the 1:1 orbital resonance with a large eccentricity
$\sim0.65$; therefore, this "Earth" has a quite large variation of $T_p$ ($\sim 220$K) and is not habitable. Only "Earths" from 1.2 AU
to 1.6 AU survive as habitable "Earths". Due to the perturbation of Kepler-47 c, these habitable "Earths" have a large $dT_p\sim$20K.
Inside the inner boundary of the HZ, there are 14 stable "Earths", with an even larger $dT_p>30$ K due to the perturbations of both
Kepler-47 b and c.

If the albedo $\alpha=0.3$, the HZ extends from 0.75 to 1.34 AU, and there will be two tolerant regions for a stable habitable
"Earth": 0.75-0.8 AU and 1.2-1.34 AU. With $\alpha=0.5$, the HZ becomes 0.64-1.13 AU, and the tolerant region becomes 0.64-0.8 AU(see
Table 2).

\section{TTVs via the Additional "Earth"}
Because all the "Earths" have an inclination=1$^\circ$, we can't observe their transits. To determine the existence of stable "Earths"
in the HZ, we can calculate the transit time ($T_{\rm tr}$) and TTVs of the existing planet in each system, which can be observed by
follow-ups. We calculate two $T_{\rm tr}$ with and without an additional "Earth" in each systems, and check their differences $dT_{\rm
tr}$. The initial conditions used here are according to our results in Section 3. The "Earths" labeled as "sample" in Figure 1 and 2,
are chosen as the sample cases. The initial conditions of both binary stars and planets in this section are the same with the
configurations in the sample cases at the end of simulations (time=10 Myr) in Section 3.

Figure 3 shows the differences of both transit time ($dT_{\rm tr}$) and TTVs ($d$TTV) in 4000 days in each system. Note that we only
show the transits of the larger binary star A in each system. For Kepler-47 system, only the transits of 47 b are considered because
the period of 47 c is so long that few transit events can be observed. In Figure 3(a), all systems have increasing variations $dT_{\rm
tr}$. In Kepler-16, the "Earth" is in the corotational resonance with Kepler-16 b, and the interaction between these two planets is
strong. Therefore, the $dT_{\rm tr}$ is more obvious than other four systems. The $dT_{\rm tr}$ of Kepler-16 b can achieve
$\sim10^{-2}$ day in 3 years and $\sim1$ day in 10 years. The other four systems have small variations due to the limited perturbation
of the additional "Earth": $dT_{\rm tr}<10^{-3}$ day in 3 years and $dT_{\rm tr}<10^{-2}$ day in 10 years. We can sort these five
systems by $dT_{\rm tr}$: Kepler-16 b$>$Kepler-34 b$\sim$Kepler-38 b$>$Kepler-35 b$>$Kepler-47 b. Figure 3(b) also shows the
calculation of the TTV of the existing planets. Similar to $dT_{\rm tr}$, the strong interaction between Kepler-16 b and co-orbit
"Earth" results in a larger $d$TTV$\sim0.1$ day. The other four systems have small $d$TTV$<10^{-2}$ day.


\section{Conclusions}
With the motivation of finding additional "Earths" in the HZs of Kepler-16, 34, 35, 38 and 47, we simulated 100 cases in each system
and show the stability and habitability of these "Earths" (see Section 3). After 10 Myr evolution, most "Earths" are unstable in
Kepler-16 and we do not find any habitable "Earth". However, two "Earths" in the corotational resonance with Kepler-16 b survived near
the outer HZ boundary when the albedo $\alpha=0$. In Kepler-47, due to the perturbation of Kepler-47 c, only eight "Earths" from 1.2
to 1.6 AU, survive and are habitable. In Kepler-34, 35 and 38, "Earths" in the HZ are stable and most "Earths" in the HZ initially can
remain in the HZ. The tolerant region of an additional habitable "Earth" in each system is shown in Table 2.

Different albedos ($\alpha=$0.3 and 0.5) of the "Earth" make the conclusions a little different. Comparing with $\alpha=0$, "Earths"
absorb less flux and have a lower $T_p$, i.e., the HZ migrates to a closer and narrower region. In the Kepler-16 system, we still do
not obtain any habitable "Earth" with $\alpha$=0.3 or 0.5. Kepler-34, 35, and 38 systems have fewer habitable "Earths" that survived
and the outer boundaries of the tolerant regions shrink. Overlapping the HZ with the unstable region in Kepler-47, there are two
tolerant regions for a habitable "Earth" in Kepler-47 when we set $\alpha=0.3$, as shown in Table 2.

Comparing the $T_{\rm tr}$ with and without the perturbation of the additional "Earth", we show that: an "Earth" in the same orbit
with Kepler-16 b can influence the $T_{\rm tr}$ of Kepler-16 b obviously, with $dT_{\rm tr}\sim1$ day and $d$TTV$\sim0.1$ day in 10
years. Meanwhile, $dT_{\rm tr}$ and $d$TTV due to the perturbation of the habitable "Earth" in other systems are relatively small.
With the precision of $10^{-3}$ day $\sim$ 88s, we can identify $d$TTV of Kepler-16 b in three years. After 10 years of observations,
the $d$TTV of Kepler-34 b and 38 b can be detected, while the $d$TTV of Kepler-35 b and 47 b are too small to be detected.



The work is supported by National Basic Research Program of China (2013CB834900) and National Natural Science Foundations of China
(Nos.10833001, 10925313,  11003010).


\begin{table}[b]
\begin{center}
\caption{Stellar and planetary characters in the five binary systems.\label{tbl-1}}
\begin{tabular}{llccccc}
\hline
         &  & Kepler-16  & Kepler-34  & Kepler-35 & Kepler-38 & Kepler-47\footnote{Data are from \citealt{Doyle11},
\citealt{Welsh12}, \citealt{Oroze12b}, \citealt{Oroze12b} and \citealt{Oroze12a} respectively.} \\
\hline
         & Mass($M_\odot$)           & 0.6897   & 1.0479    & 0.8877    & 0.949          & 1.043 \\
Star A   & Radius($R_\odot$)         & 0.6489   & 1.1618    & 1.0284    & 1.757          & 0.964 \\
         & Temperature(K)            & 4450     & 5913      & 5606      & 5623           & 5636  \\
\hline
         & Mass($M_\odot$)           & 0.20255  & 1.0208    & 0.8094    & 0.249          & 0.362 \\
Star B   & Radius($R_\odot$)         & 0.22623  & 1.0927    & 0.7861    & 0.2724         & 0.3506\\
         & Temperature(K)            & 2800     & 5867      & 5202      & 3315           & 3357  \\
\hline
         & semi-major axis(AU)       & 0.22431  & 0.2288    & 0.17617   & 0.1469         & 0.0836\\
binary   & period(days)              & 41.079220& 27.7958103& 20.733667 & 18.79537       & 7.44837695\\
parameters & eccentricity            & 0.15944  & 0.52      & 0.1421    & 0.1032         & 0.0234\\
         & $T_{\rm eff}$(K)\footnote{$T_{\rm eff}$ is the effective temperature of the binary system}          & 4307.8   & 5892.0    & 5435.5    & 5496.7         & 5501.5\\
\hline
         & Mass($M_{\rm J}$)         & 0.333    & 0.22      & 0.127     & 0.3964$R_{J}$  & 0.27$R_{\rm J}$  \\
         & semi-major axis(AU)       & 0.7048   & 1.0896    & 0.603     & 0.4644         & 0.2956\\
Planet b & period(days)              & 228.776  & 288.822   & 131.455   & 105.595        & 49.514\\
         & eccentricity              & 0.0069   & 0.182     & 0.042     & 0.016          & $<0.035$\\
         & $T_p$($\alpha=0$)\footnote{$T_p$ is the equilibrium temperature of planet with albedo $\alpha=0$.}                    & 200-215  & 310-390   & 370-420   & 460-490        & 480-510\\
\hline
         & Mass($M_{\rm J}$)         &    -     &   -       &    -      &     -          & 0.42$R_{\rm J}$  \\
         & semi-major axis(AU)       &    -     &   -       &    -      &     -          & 0.989 \\
Planet c & period(days)              &    -     &   -       &    -      &     -          & 303.158\\
         & eccentricity              &    -     &   -       &    -      &     -          & $<0.411$\\
         & $T_p$($\alpha=0$)                    &    -     &   -       &    -      &     -          & 267-273\\

\hline
\end{tabular}
\end{center}
\end{table}

\begin{table}
\begin{center}
\caption{Main results in the five binary systems. \label{tbl-1}}.
\begin{tabular}{lccccc}
\hline
         & Kepler-16  & Kepler-34  & Kepler-35 & Kepler-38 & Kepler-47 \\
\hline
Stable "Earth"  & 7          & 74        &  85        & 100         & 66\\
Stable "Earth" in HZ              & $\leq$2       & 32        &  36        & 38          & 7\\
\hline
Stable region (AU)           & $>$0.66    & $>$1.5    & $>$1       & $>$1.1      & 0.4-0.8 , $>$1.2   \\
Tolerant region (AU)     & & & & & \\
~~$\alpha$=0              & $\sim$0.72 & 1.6-2.8   & 1.2-2.0    &  1.6-2.9    &  1.2-1.6  \\
~~$\alpha$=0.3            & Null       & 1.6-2.34  & 1.0-1.67   &  1.34-2.43  &  0.75-0.8 , 1.2-1.34   \\
~~$\alpha$=0.5            & Null       & 1.6-1.98  & 0.85-1.41  &  1.13-2.05  &  0.64-0.8  \\
\hline
$dT_p$(K)\footnote{Variation of $T_p$ with albedo $\alpha=0$}                   & $<20$  & $<12$ & $<8$ & $<10$  &  $<30$\\
$dT_{\rm tr}$(days)\footnote{Difference between transit times $T_{\rm tr}$ in 10 years}        & 1 & $2\times10^{-2}$ & $2\times10^{-3}$ & $10^{-2}$ & $3\times10^{-4}$ \\
$dTTV$(days)\footnote{Differences between TTVs in 10 years}       & 0.1 & $10^{-2}$ & $10^{-3}$ & $3\times10^{-3}$ &  $10^{-4}$   \\

\hline
\end{tabular}
\end{center}
\end{table}

\begin{figure}
\vspace{0cm}\hspace{0cm} \epsscale{1.0} \plotone{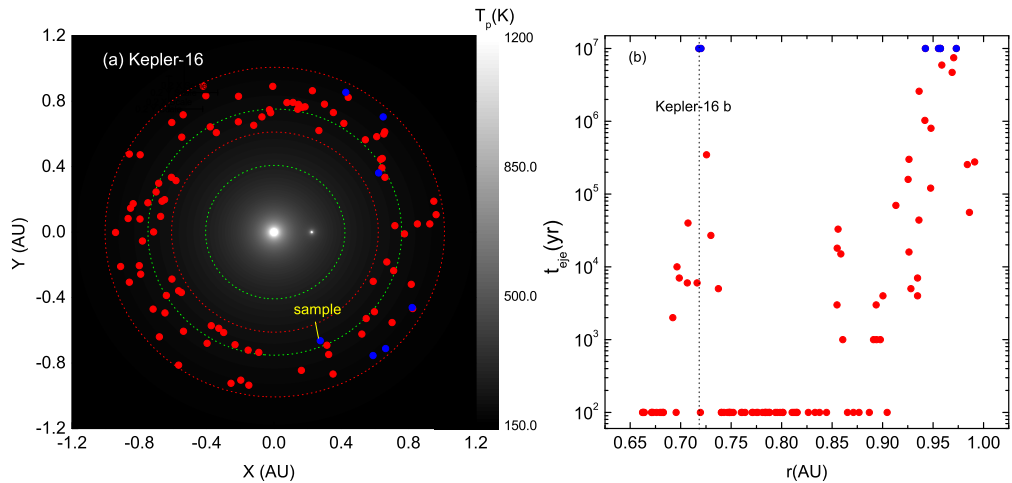} \vspace{0cm} \caption{"Earths" in Kepler-16: (a) two green dotted lines are
the HZ boundaries with albedo $\alpha=0$. Temperatures of each line are labeled. The two red dotted lines represent the extension of
our 100 "Earths",  with initial random locations shown in the plot. The red/blue circles represent unstable/stable "Earths" after 10
Myr evolution. The one labeled as "sample" is chosen for the case in Section 4. (b) The ejected time of the 100 "Earths". $r$ is the
distance between the "Earths" and Kepler-16 A (and hereafter). The dotted line represents the location of Kepler-16 b.  \label{fig1}}
\end{figure}

\begin{figure}
\vspace{0cm}\hspace{0cm} \epsscale{0.8} \plotone{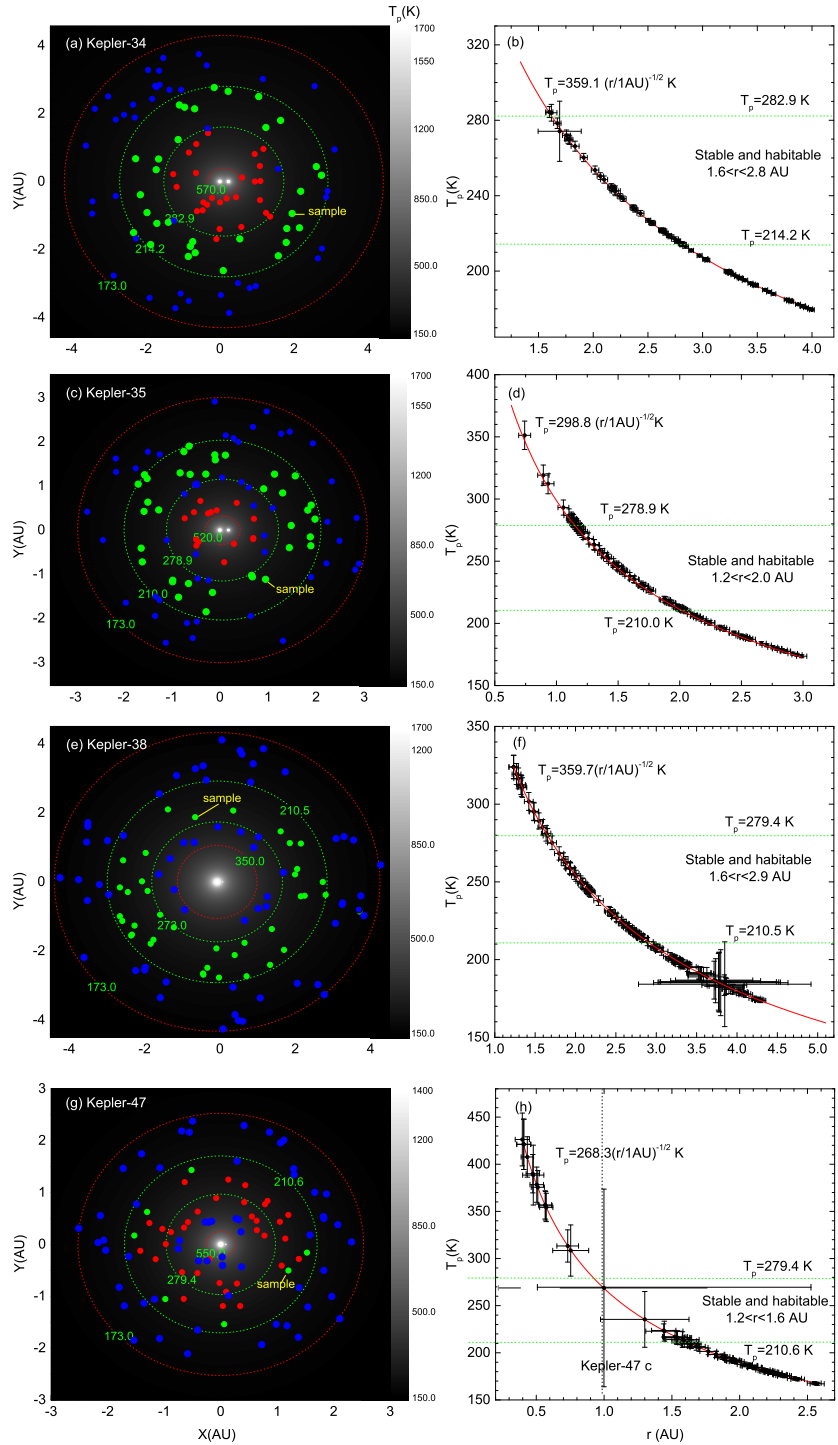} \vspace{0cm} \caption{"Earths" in Kepler-34, 35, 38 and 47 are represented
in panels (a),(c),(e) and (g), respectively: The green filled circles represent stable habitable "Earths", while the red/blue circles
stand for unstable/stable "Earths" outside the HZ after 10 Myr. The green and red dotted lines are the boundaries of the HZ and the
initial locations of 100 "Earths", respectively. The equilibrium temperature $T_p$ and the location $r$ of stable "Earths" in
Kepler-34, 35, 38 and 47 are shown in panels (b),(d),(f) and (h), respectively. The error bars show the largest and smallest values of
$T_p$ and $r$. \label{fig2}}
\end{figure}

\begin{figure}
\vspace{0cm}\hspace{0cm} \epsscale{0.8} \plotone{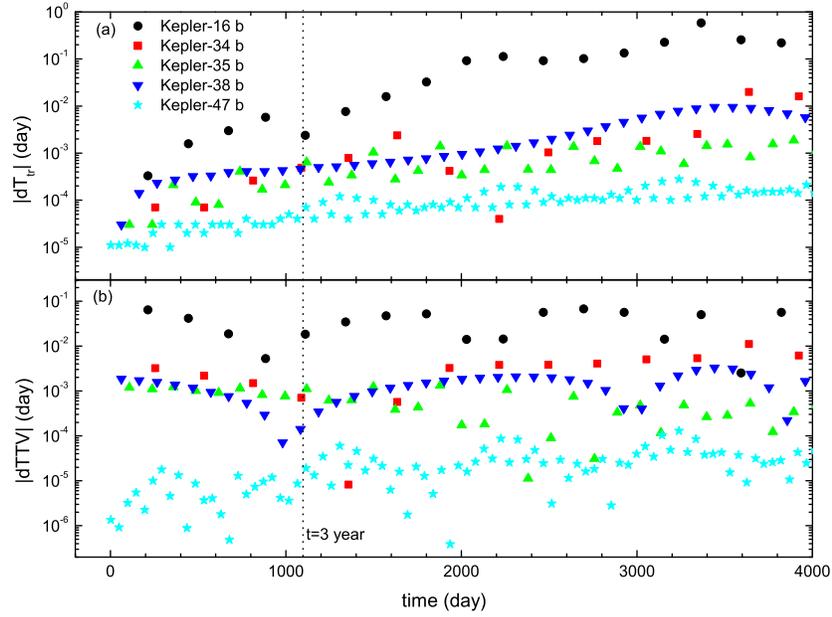} \vspace{0cm} \caption{Differences of (a) transit times ($dT_{\rm tr}$) and
(b) TTV($d$TTV) of detected planets with and without the perturbation of an additional "Earth". Because of the stronger perturbation
between "Earth" and Kepler-16b in 1:1 orbital resonance, $T_{\rm tr}$ and $d$TTV of Kepler-16b are obviously larger than other four
systems, and can be detected in three years. \label{fig3}}
\end{figure}

\end{document}